      \documentclass[preprint2]{aastex}

\shorttitle{A new contact binary: 2002~CC$_{249}$?}
\shortauthors{Audrey Thirouin, and Scott. S. Sheppard}

\begin{document}

\title{A possible dynamically Cold Classical contact binary: (126719) 2002~CC$_{249}$.}

\author{Audrey Thirouin\altaffilmark{1}}
\affil{Lowell Observatory, 1400 W Mars Hill Rd, Flagstaff, Arizona, 86001, USA. }
\email{thirouin@lowell.edu}

\and
 
\author{Scott S. Sheppard\altaffilmark{2}}
\affil{Department of Terrestrial Magnetism (DTM), Carnegie Institution for Science, 5241 Broad Branch Rd. NW, Washington, District of Columbia, 20015, USA.}
 \email{ssheppard@carnegiescience.edu}

\begin{abstract} 

Images of the Kuiper belt object (126719) 2002~CC$_{249}$ obtained in 2016 and 2017 using the 6.5m Magellan-Baade Telescope and the 4.3m Discovery Channel Telescope are presented. A lightcurve with a periodicity of 11.87$\pm$0.01~h and a peak-to-peak amplitude of 0.79$\pm$0.04~mag is reported. This high amplitude double-peaked lightcurve can be due to a single elongated body, but it is best explained by a contact binary system from its U-/V-shaped lightcurve. We present a simple full-width-at-half-maximum (FWHM) test that can be used to determine if an object is likely a contact binary or an elongated object based on its lightcurve. Considering that 2002~CC$_{249}$ is in hydrostatic equilibrium, a system with a mass ratio q$_{min}$=0.6, and a density $\rho$$_{min}$=1~g~cm$^{−3}$, or less plausible a system with q$_{max}$=1, and $\rho$$_{max}$=5~g~cm$^{−3}$ can interpret the lightcurve. Assuming a single Jacobi ellipsoid in hydrostatic equilibrium, and an equatorial view, we estimate $\rho$$\geq$0.34~g~cm$^{-3}$, and a/b=2.07. Finally, we report a new color study showing that 2002~CC$_{249}$ displays an ultra red surface characteristic of a dynamically Cold Classical trans-Neptunian object.  

\end{abstract}


\section{Introduction}

The trans-Neptunian (or Kuiper) belt is structured in four dynamical groups: i) \textit{classical trans-Neptunian Objects (TNOs)} are between 40 and 48~AU, and are not significantly perturbed by Neptune or captured in a mean motion resonance with Neptune. Their orbits have low inclinations and are almost circular (typically with eccentricity $<$0.3, ii) \textit{resonant TNOs} are trapped in a resonance with Neptune and thus have had significant interactions with Neptune in the past, iii) \textit{scattered disk TNOs} have large inclinations and eccentricities, with perihelia near Neptune's orbit, suggesting they were scattered by Neptune in the past and iv) \textit{extreme or detached TNOs} with highly eccentric orbits present perihelion distances (q$>$40AU) beyond the Neptune gravitational influence. 

Based on orbital inclination, size and color studies of objects in the classical belt, at least two sub-populations have been identified \citep{Peixinho2008, Brown2001, Levison2001}: i) the Hot classical TNOs are dynamically excited, have high orbital inclination and eccentricity and were likely scattered by the giant planets and captured into the trans-Neptunian population, ii) the Cold Classical TNOs at low inclinations appear more primordial, are small, and are red \citep{Batygin2011, Benecchi2009, Noll2008book}.

Among the Cold Classical population, the separated binary fraction is high, about 20-25$\%$ but in the other dynamical groups the percentage is only 5-10$\%$ \citep{Noll2008book}. Most of the binary Cold Classicals are wide equal-sized binaries with primary and secondary having comparable sizes. In the trans-Neptunian belt, the contact binary population remains elusive. The first and unique confirmed contact binary is 2001~QG$_{298}$ (an object in the 3:2 mean motion resonance with Neptune) whereas 2003~SQ$_{317}$ (a Haumea family member, dynamically Hot Classical), and 2004~TT$_{357}$ (an object in the 5:2 mean motion resonance with Neptune) are likely contact binaries \citep{Sheppard2004, Lacerda2014, Thirouin2017}. Surprisingly, none of these objects are in the dynamically Cold Classical sub-population, where the highest fraction of wide binaries is seen.  

2002~CC$_{249}$ is a dynamical Cold Classical TNO with a semi-major axis\footnote{Orbital elements from the Minor Planet Center, August 2017.} of 47.44~AU, an inclination of 0.84$^\circ$, and an eccentricity of 0.20 \citep{Gladman2008}. Based on \textit{Hubble Space Telescope} images, no companion orbiting 2002~CC$_{249}$ was detected \citep{Noll2008}. 

Following, we present the lightcurve of 2002~CC$_{249}$ based on observations carried out since 2016. The lightcurve has a large variability caused by an egg-shaped object or more likely by a binary system in close configuration. We also report new color estimates for this object. Next, we describe our observations, and the data reduction techniques. Sections 3 and 4 present and analyze the lightcurve and the colors of 2002~CC$_{249}$. Finally, we summarize our results in the latest section of this paper.


\section{Observations and analysis}
 
We present in-situ observations carried out with the Discovery Channel Telescope (DCT) and the 6.5~m Magellan Telescope (Baade unit) in 2016 and in 2017. 

The DCT is located in Arizona (Happy Jack, United States of America). Our observations were obtained with the Large Monolithic Imager (LMI). This instrument is a 6144$\times$6160 pixels CCD for a total field of view of 12.5$\arcmin$$\times$12.5$\arcmin$, and a pixel scale of 0.12$\arcsec$/pixel \citep{Levine2012}.  

At the Las Campanas Observatory in Chile, we used one of the Magellan Twin Telescopes. The Inamori-Magellan Areal Camera \& Spectrograph (IMACS) mounted on the Baade telescope is a wide-field imager with eight 2048$\times$4096~pixels CCDs. The short camera mode was selected for a pixel scale of 0.20$\arcsec$/pixel and a 27.4$\arcmin$ diameter field. 

The lightcurve study was performed at DCT using the VR-filter or r'-filter. The color study was performed with the g'r'i' Sloan filters at the Magellan-Baade telescope, and DCT. Our basic observing log is reported in Table~\ref{Tab:Log_Obs}. The calibration and reduction of our images were performed following the procedure described in \citet{Thirouin2014, Thirouin2016}. The search for periodicities has been done with the same techniques mentioned in \citet{Thirouin2017}. Finally, the color and solar phase curve studies were performed as in \citet{Thirouin2012}. For the color study, the best fit aperture radius varied between 4 and 5 pixels for the Magellan-Baade data and was about 3 pixels for the DCT data.


\section{Photometric results}
\label{sec:lightcurve}

\subsection{Color and solar phase curve}

\citet{SantosSanz2009} observed 2002~CC$_{249}$ for color studies with the Very Large Telescope\footnote{\citet{SantosSanz2009} used the Antu unit at the Very Large Telescope (ESO-VLT, Cerro Paranal, Chile). They used the FORS1 detector and Bessel broadband BVRI filters. Details can be found in \citet{SantosSanz2009}} on March 25, 2004 (UT) at a phase angle $\alpha$=0.4$^\circ$. They calculated: \textit{V-R}=0.51$\pm$0.08~mag, \textit{R-I}=0.69$\pm$0.06~mag, \textit{V-I}=1.20$\pm$0.07~mag, and found a spectral gradient of 22.3$\pm$8.5$\%/100~nm$. They also derived the absolute magnitudes in the R and V bands using the Bowell formalism and the linear formalism \citep{SantosSanz2009}: H$_{V}(linear)$=6.50$\pm$0.06 mag, and H$_{R}(linear)$ = 5.99$\pm$0.05 mag, and found a R-magnitude of 21.87$\pm$0.05~mag. It is important to point out that \citet{SantosSanz2009} did not know the lightcurve of 2002~CC$_{249}$, and thus they have not removed the brightness variation due to rotation for their color estimates. Their data have been obtained over about 20~min. With such a short duration\footnote{No Julian Date are available in \citet{SantosSanz2009}, but the data are available in the European Southern Observatory (ESO) archive system at: \url{http://archive.eso.org/cms.html}.}, they were not able to notice the large amplitude, and slow rotation of 2002~CC$_{249}$.   

Most of our data have been obtained with a VR-broadband filter, and thus are not ideal for color study nor solar phase curve study. However, we have two sets of color data\footnote{Data obtained on March 10, 2017 are not considered for the color/phase curve study as the weather conditions were not photometric.} suitable for these studies. Unfortunately, the g' and i' bands observed at DCT have an insufficient quality to be included here, only the r' band will be used for the solar phase curve of 2002~CC${249}$. 
The phase function is: 
\begin{equation}
\phi (\alpha) = 10^{-0.4 \alpha\beta} 
\end{equation}
with $\alpha$ as phase angle and $\beta$ as the phase coefficient at $\alpha$$<$2$^\circ$.
Based our Magellan and DCT data, the range of phase angles is limited with observations of 2002~CC$_{249}$ at 0.6$^\circ$ and 1$^\circ$ for color study. However, by including \citet{SantosSanz2009} data, the solar phase curve of 2002~CC$_{249}$ is over a phase angle range between 0.4$^\circ$ and 1$^\circ$. 
Based on \citet{Smith2002, Sheppard2012}, once can converted the Johnson-Morgan-Cousins colors (BVRI, used by \citet{SantosSanz2009}) to the Sloan colors (g'r'i'z'): V-R=0.59(g'-r')+0.11, and R-I=1.00(r'-i')+0.21.Using previous equations, \citet{SantosSanz2009} obtained g'-r'=0.68~mag, and r'-i'=0.48~mag.

\begin{figure*}
\includegraphics[width=9cm, angle=0]{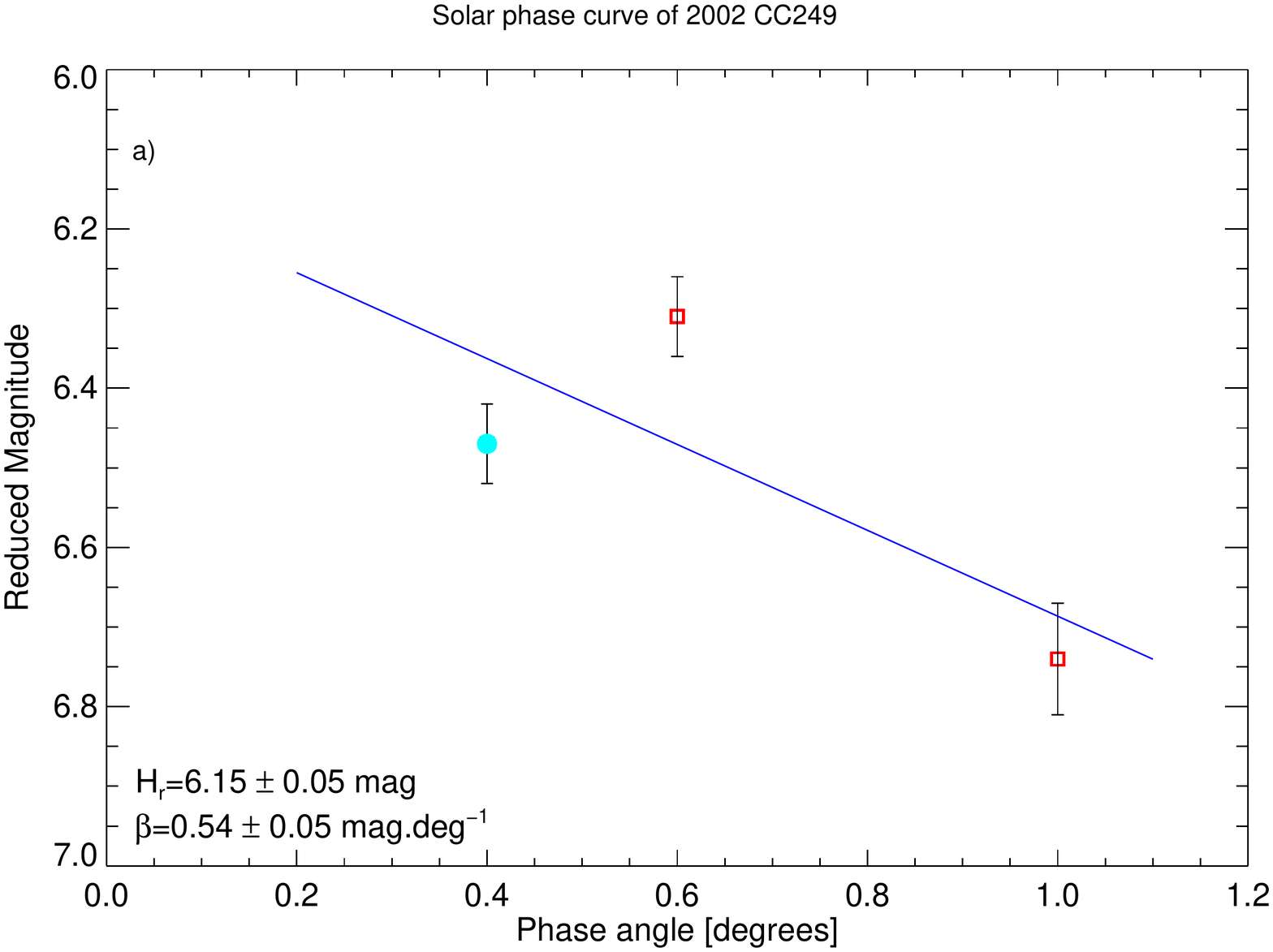}
\includegraphics[width=9cm, angle=0]{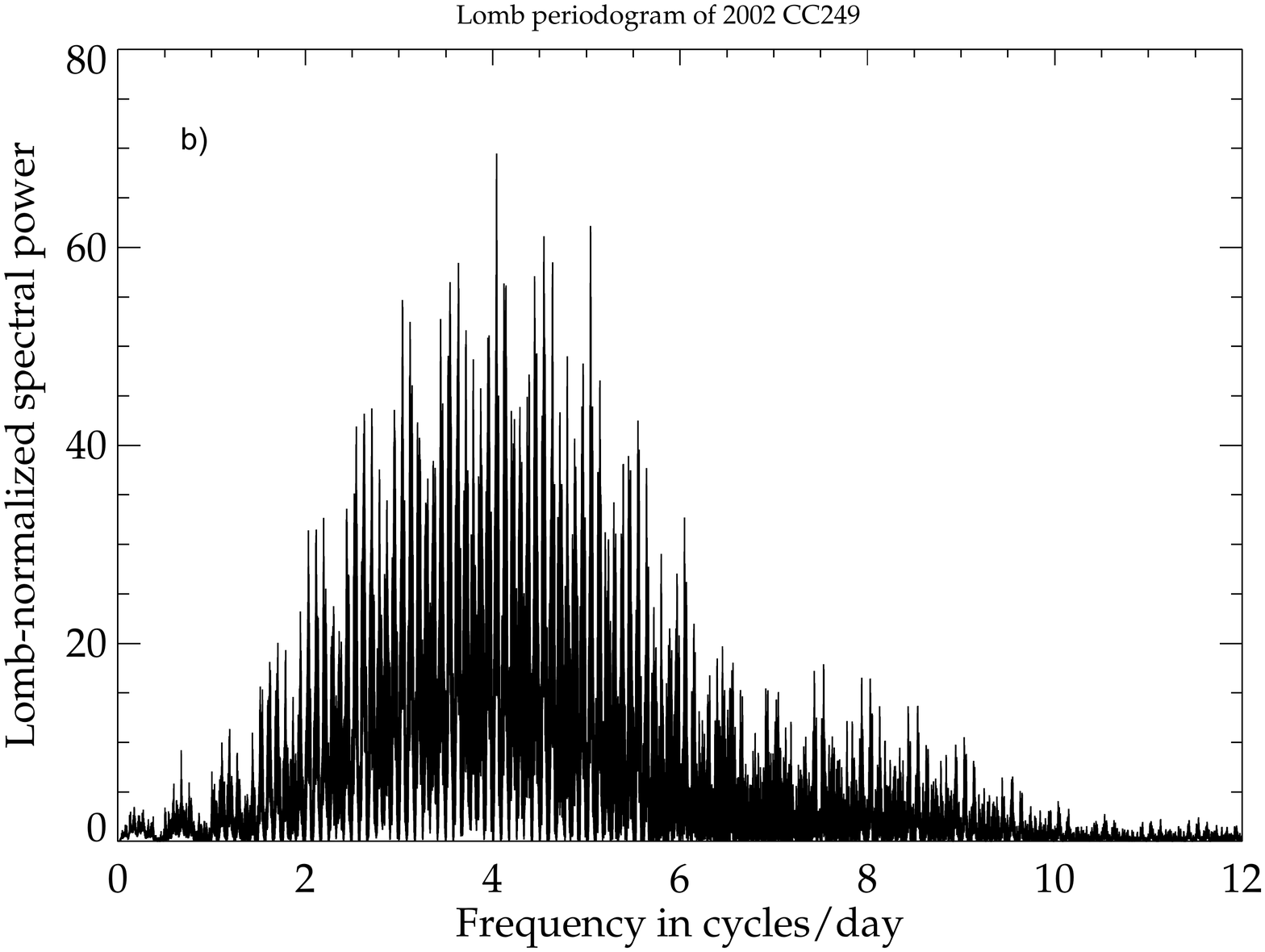}
\includegraphics[width=9cm, angle=0]{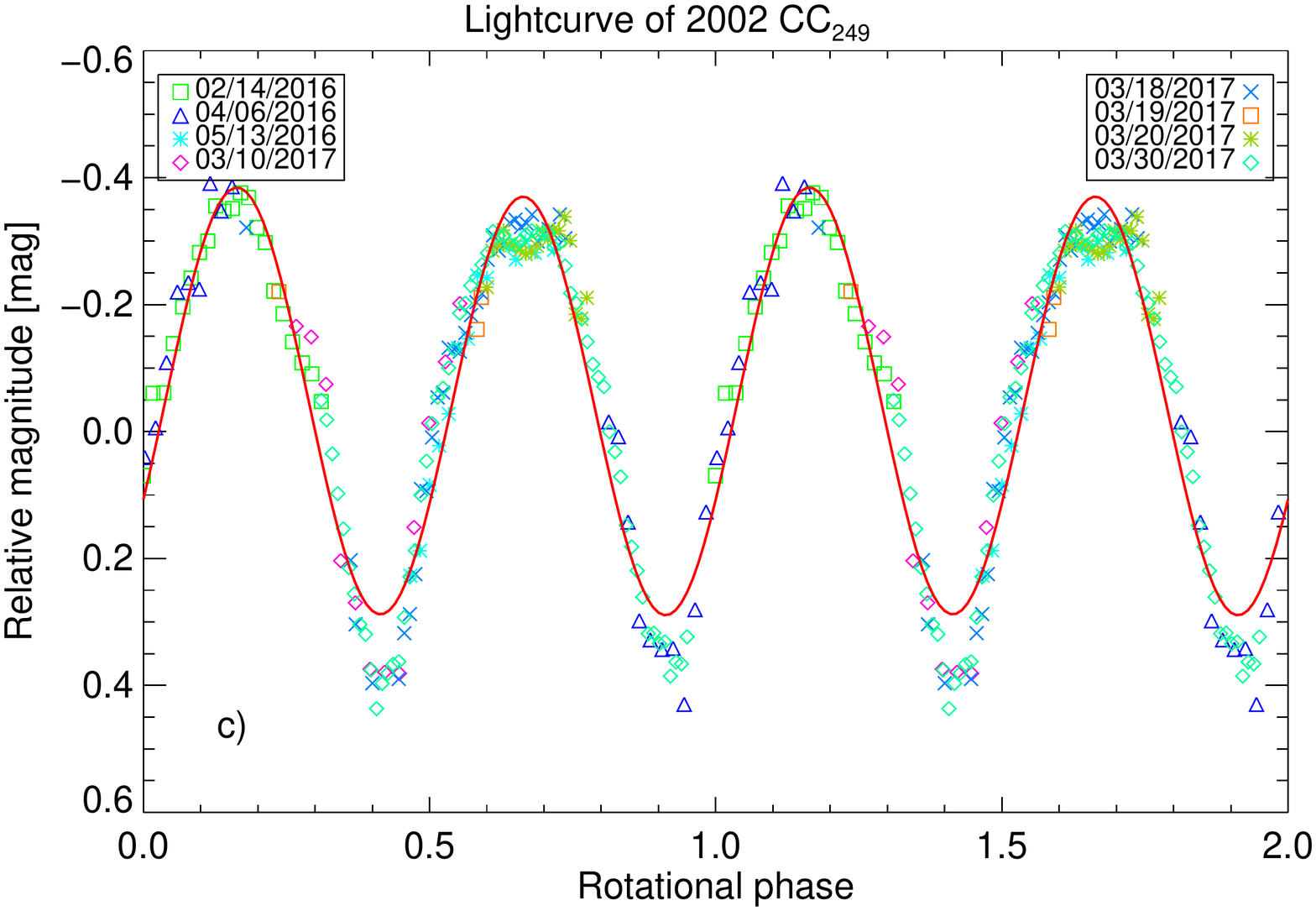}
\caption{\textit{Solar Phase curve, Lomb periodogram, and lightcurve of 2002~CC$_{249}$}: The solar phase curve (a)) is plotted using our data (red square), and \citet{SantosSanz2009} result (cyan circle). The peak with the highest spectral power of the Lomb periodogram is located at 4.04~cycles/day (b)). The double-peaked lightcurve is plotted over two rotations. The 2$^{nd}$ order Fourier series (red continuous line) is not able to reproduce the V-and U-shape of the curve (c)). Error bars are not plotted for clarity, but the typical error bar is $\pm$0.05~mag for the photometry}.
\label{fig:LCLomb}
\end{figure*}

The absolute magnitude (H$_{r'}$) is the object's magnitude assuming that the object is at 1~AU from the Sun (r$_{h}$) and the Earth ($\Delta$) and at $\alpha$=0$^\circ$: 
\begin{equation}
H_{r'}= m_{r'} (1,1,\alpha=0^\circ) = m_{r'} -5 \log (r_h \Delta) -\alpha\beta
\end{equation}
where the corrected r'-band magnitude is m$_{r'}$(1,1,$\alpha$). With brightness variations due to rotation, and the distance removed, we obtain: $\beta$~=~0.54$\pm$0.05~mag/$^\circ$, H$_{r'}$~=~6.15$\pm$0.05~mag (Figure~\ref{fig:LCLomb}). However, the value from \citet{SantosSanz2009} is not corrected from brightness variation as the rotational phase for this point cannot be estimated securely. In fact, even if we are able to predict the rotational phase of the \citet{SantosSanz2009} data, the propagation of the uncertainty for the rotational period estimate is to be considered. On the other hand, because these two datasets are separated by more than 13 years, we may also have to consider that the lightcurve have changed over the years due to change in the system geometry (or pole orientation if it is a single object). Therefore, there is an uncertainty of $\pm$0.4~mag for \citet{SantosSanz2009} data due to the brightness variation of the object (error bar due to brightness not plotted in Figure~\ref{fig:LCLomb} for clarity). In conclusion, our phase curve is not optimal and more data are required to provide a clear and secure solar phase curve.

We also use our Magellan dataset for color study and report: g'-i'=1.24$\pm$0.05~mag, g'-r'=0.97$\pm$0.06~mag, and r'-i'=0.27$\pm$0.06~mag. In conclusion, 2002~CC$_{249}$ displays an ultra red surface characteristic of a dynamically cold classical TNO based on the \citet{SantosSanz2009} study and this work.

\subsection{Lightcurve}

Our dataset is composed of three isolated nights in 2016 as well as two isolated and three consecutive nights in 2017. During our observations in 2016, only fragments of the lightcurve of 2002~CC$_{249}$ were obtained. Our longest run was $\sim$3.7~h, and only one maximum of the curve with an amplitude of about 0.5~mag was observed. Therefore, a long rotational period (P$>$8~h, assuming a double-peaked lightcurve) was suspected. One maximum and one minimum were observed on UT March 18, 2017, and two minima and one maxima on UT March 30. Both nights allowed us to constrain the rotational period to approximately 12~h assuming a double-peaked lightcurve.

We applied a light-time correction to our observing runs. The highest peak of the Lomb periodogram is at 4.04~cycles/day (5.94~h), and the PDM method confirms such a peak (Figure~\ref{fig:LCLomb}). The next step is to select the best option between single- and double-peaked lightcurve (i.e. period of 5.94~h or 11.87~h). 

Generally, the albedo contributions is up to 20$\%$ for asteroids and TNOs \citep{Degewij1979, Sheppard2008, Thirouin2010}. Some TNOs like Eris, Makemake, or Haumea have high geometric albedos between 51\% and 96\% \citep{Sicardy2011, Ortiz2012, Ortiz2017}. The lightcurves of Eris and Makemake are mostly flat because dominated by the nearly spherical shape and/or pole-on orientation of these objects, whereas in the case of Haumea, the lightcurve is dominated by Haumea's elongated shape, and the dark red spot contribution is only about 10\% \citep{Thirouin2013, Lacerda2008}. Assuming that 2002~CC$_{249}$ has a single-peaked lightcurve, albedo variation(s) of about 80$\%$ would be required on the object's surface. This scenario is unlikely, and therefore the single-peaked option seems inadequate. Secondly, by plotting the double-peaked lightcurve, once can appreciate that there is a $\sim$0.1~mag asymmetry between the first and second maxima. In conclusion, the double-peaked option is the more adequate for 2002~CC$_{249}$. The double-peaked lightcurve assuming a periodicity of 11.87~h and a full amplitude of 0.79$\pm$0.04~mag is plotted over two cycles in Figure~\ref{fig:LCLomb}. In Table~\ref{Tab:Summary_photo}, we report the photometry used in this work. The zero phase of the lightcurve is the date of the object's first image (Table~\ref{Tab:Summary_photo}).  

With such a large lightcurve amplitude, 2002~CC$_{249}$ can be a contact binary system with a non-equator-on configuration assuming two objects with similar sizes or a single very elongated object close to an equator-on configuration (see Section~\ref{sec:analysis} for more details). A lightcurve with a U-/V-shape at the maximum/minimum of brightness is characteristic of a contact/close binary system with a near equator-on orientation \citep{Sheppard2004, Lacerda2011}. For 2002~CC$_{249}$, one can note the V-shape at the minima and the second maximum with a U-shape. But, the first maximum displays a sharper peak, and thus the U-shape is not obvious. However, it is important to point out that the first maximum is based on fragmentary datasets obtained in 2016. In Section~\ref{sec:UVshapes}, we will discuss the ``definition" of the U- and V-shapes.


\section{Analysis}

\label{sec:analysis}

\subsection{V-shape and U-shape: definition}
\label{sec:UVshapes}
 
Hektor, the largest Jupiter Trojan asteroid, was found to have large amplitude short-term light variations with a characteristic U/V shaped lightcurve \citep{Cook1971}. This U/V shaped rotational lightcurve results from shadowing and viewing geometry effects from a contact binary viewed nearly equator-on \citep{Hartmann1978, Wijesinghe1979, Weidenschilling1980}. The U/V shape of a rotational lightcurve for a contact binary asteroid based on viewing geometry and phase angle effects has been further modeled in detail by several authors \citep{Cellino1989, Lacerda2007, Gnat2010, Descamps2015}. The U-shape for the maximum and V-shape for the minimum peak of a lightcurve is apparent for contact binary asteroids viewed near equator on and at low phase angles, for which the latter occurs for all TNOs.

We here present a simple way to determine if a rotational lightcurve is likely caused by a contact binary object based on the differences in the full-width-at-half-maximum (FWMH) for the maximum (U-shape) and minimum (V-shape) peaks of brightness for the lightcurve. This analysis allows a more quantitative approach than simple visual inspection of a lightcurve without requiring a detailed model of the lightcurve. This analysis is based on the fact that the U-shape maxima of the lightcurve should show a higher FWHM than the V-shape minimum of the lightcurve, if there are differences in the peaks and it is caused by a contact binary. The full peak-to-peak amplitude has been used to estimate the U- and V-FWHM.
 
\begin{figure*}
\includegraphics[width=15cm, angle=0]{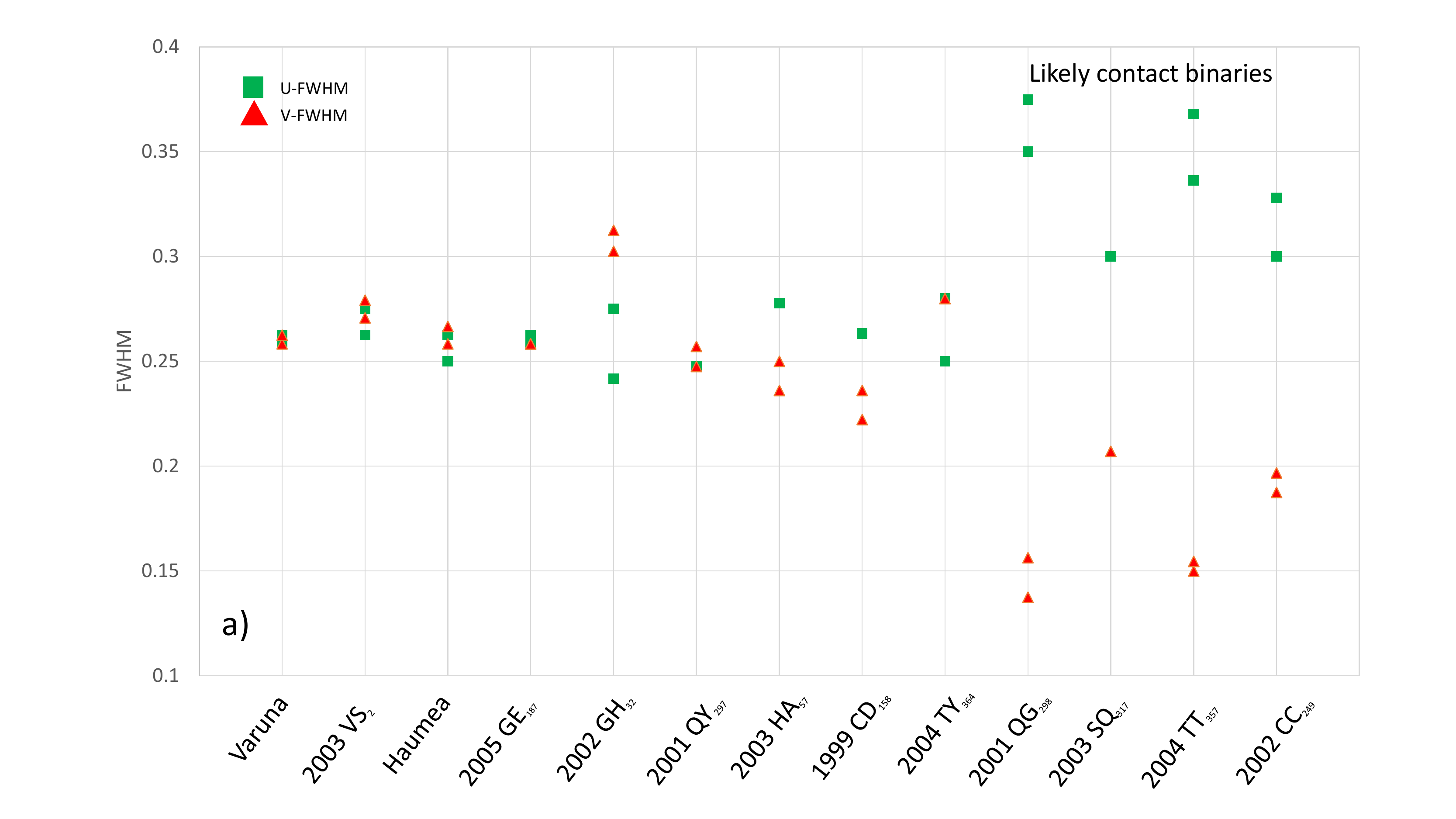}
\includegraphics[width=12cm, angle=0]{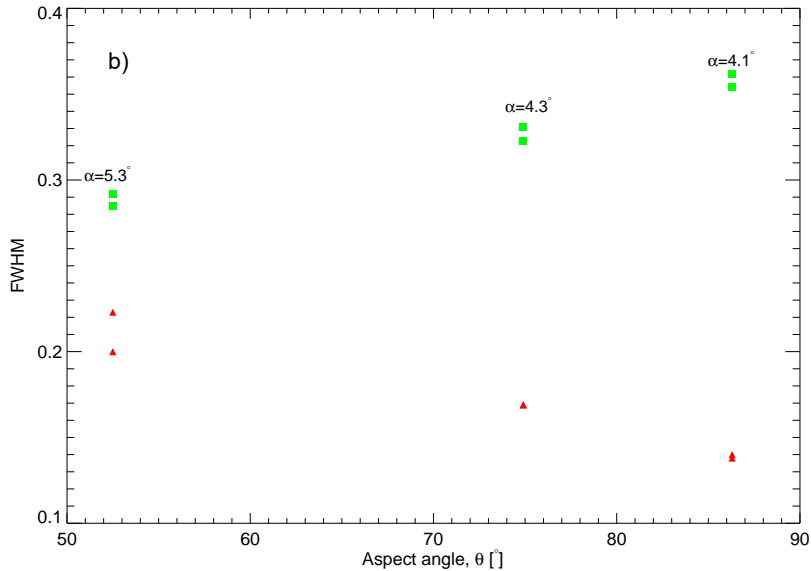}
\caption{\textit{Full width at half maximum (FWHM) of single objects, resolved binaries and (likely/confirmed) contact binaries}. \textit{Plot a)}: The FWHM of the maxima (U-FWHM) and of the minima (V-FWHM) of several objects are reported. Only double-peaked lightcurve with a amplitude larger than 0.15~mag are considered. We report the FWHM of both peaks (4 points per object), but in some cases both peaks have the same FWHM. The non-contact binary objects have a U-FWHM$\leqslant$0.28 whereas the (likely/confirmed) contact binaries present a U-FWHM$\geqslant$0.30. The V-FWHM is $\leqslant$0.21 for the (likely/confirmed) contact binaries. The minima and maxima FWHM peak differences are greater than about 0.1 for contact binaries and less than 0.05 for other objects. 2001~QG$_{298}$ is the only confirmed contact binary \citep{Sheppard2004}. \textit{Plot b)}: The U-/V-FWHM of the Jupiter Trojan Hektor versus the aspect angle of the system are plotted. Phase angles ($\alpha$) are also indicated for each dataset. One can appreciate that the FWHM differences are greater than 0.1 at large aspect angle, and is about 0.1 for an aspect angle around 53$^{\circ}$. Same legend for both plot.}
\label{fig:FWHM}
\end{figure*}

We show the FWHM of TNOs with large amplitude lightcurves ($\Delta m$$>$0.15~mag) from \citet{Thirouin2010, Thirouin2012, Thirouin2014, Thirouin2016, Sheppard2007, Jewitt2002} and likely contact binaries from \citet{Thirouin2017, Lacerda2014, Sheppard2004} in Figure~\ref{fig:FWHM}. In this Figure, we plot all four FWHM of these objects: two for the U-FWHM for the maximum and two for the V-FWHM for the minimum. It is clear the non-contact binaries or single objects have all of their peaks FWHM less than about 0.30 and near each other whereas the likely contact binaries have U-FWHM greater than about 0.30 and V-FWHM less than about 0.20. The differences between the two types of peaks is usually greater than about 0.1 for their FWHM for the contact binaries, whereas for the non-contact binaries the differences between the various peaks is less than about 0.05. 

It appears the maximum and minimum peaks of single objects have similar FWHM peaks throughout the lightcurve, whereas the likely contact binaries have significantly different FWHM for their maximum and minimum peaks.
In Figure~\ref{fig:FWHM}, we also report the evolution of the V- and U-FWHM of Hektor with the aspect angle of the system (based on \citet{Lacerda2007}). One can appreciate that at high aspect angle, the differences between the two types of peaks is above 0.1, whereas at an aspect angle of 53$^\circ$, the difference is near 0.1. For lower aspect angles, the difference is less pronounced.  

Again, this is just a simple way to quickly assess if an object's lightcurve displays a contact binary nature. A full model of the objects likely shape and configuration is needed to fully analyze an objects lightcurve. In conclusion, we consider that the U/V shape at the maximum/minimum of brightness are significantly different for likely contact binary objects and can be quantitatively looked at by the difference in their FWHM. We find that 2002~CC$_{249}$ has over a 0.1 difference in its U-FWHM versus its V-FWHM, signifying it is likely a contact binary like 2001~QG$_{298}$, 2004~TT$_{357}$ and possibly 2003~SQ$_{317}$ (Figure~\ref{fig:FWHM}). 

 \subsection{Roche system}

The large variability of 2002~CC$_{249}$ and its U-/V-shaped lightcurve is best explained if this object is a contact binary.  

\begin{figure*}
\includegraphics[width=12cm, angle=0]{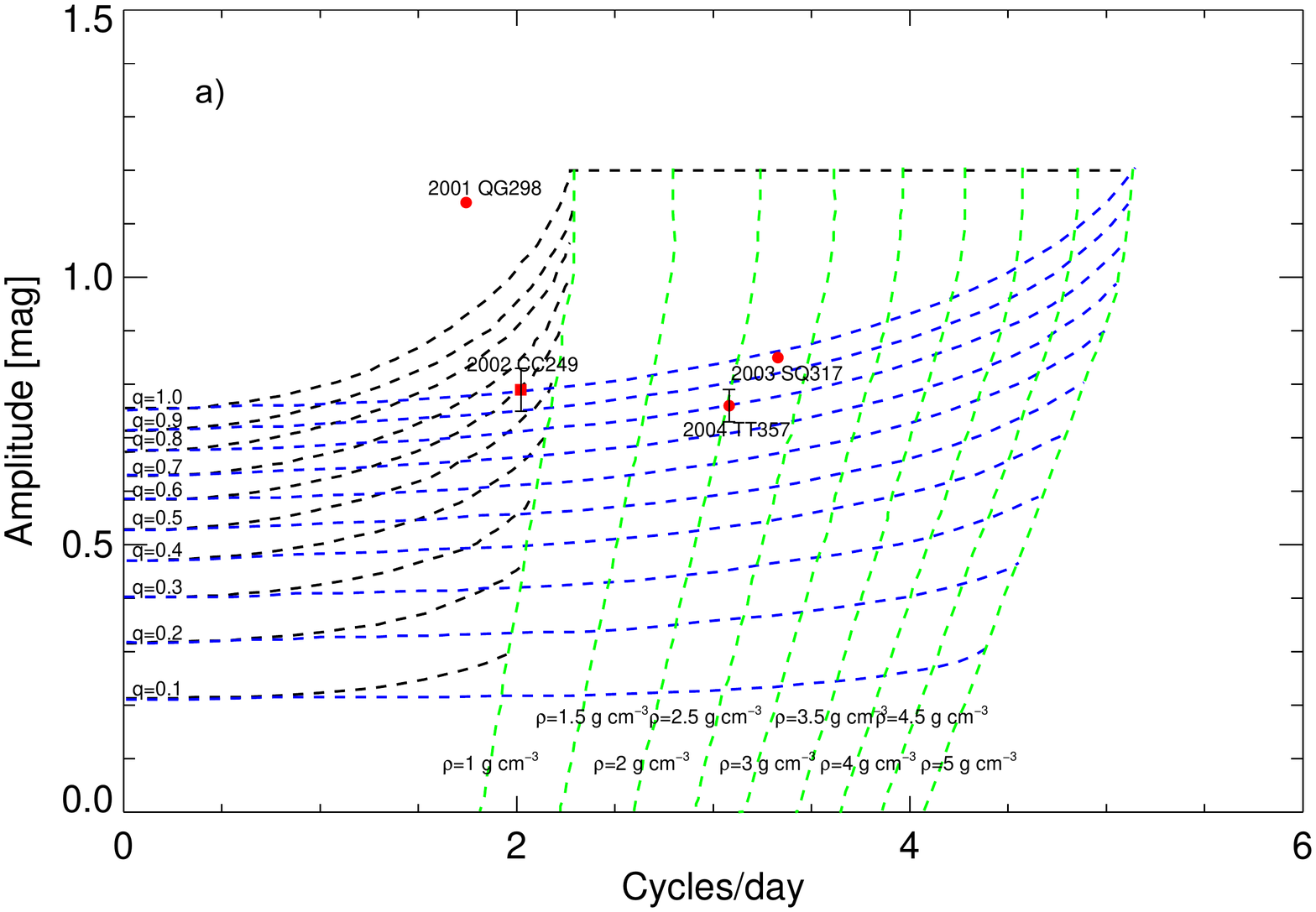}
\includegraphics[width=12cm, angle=0]{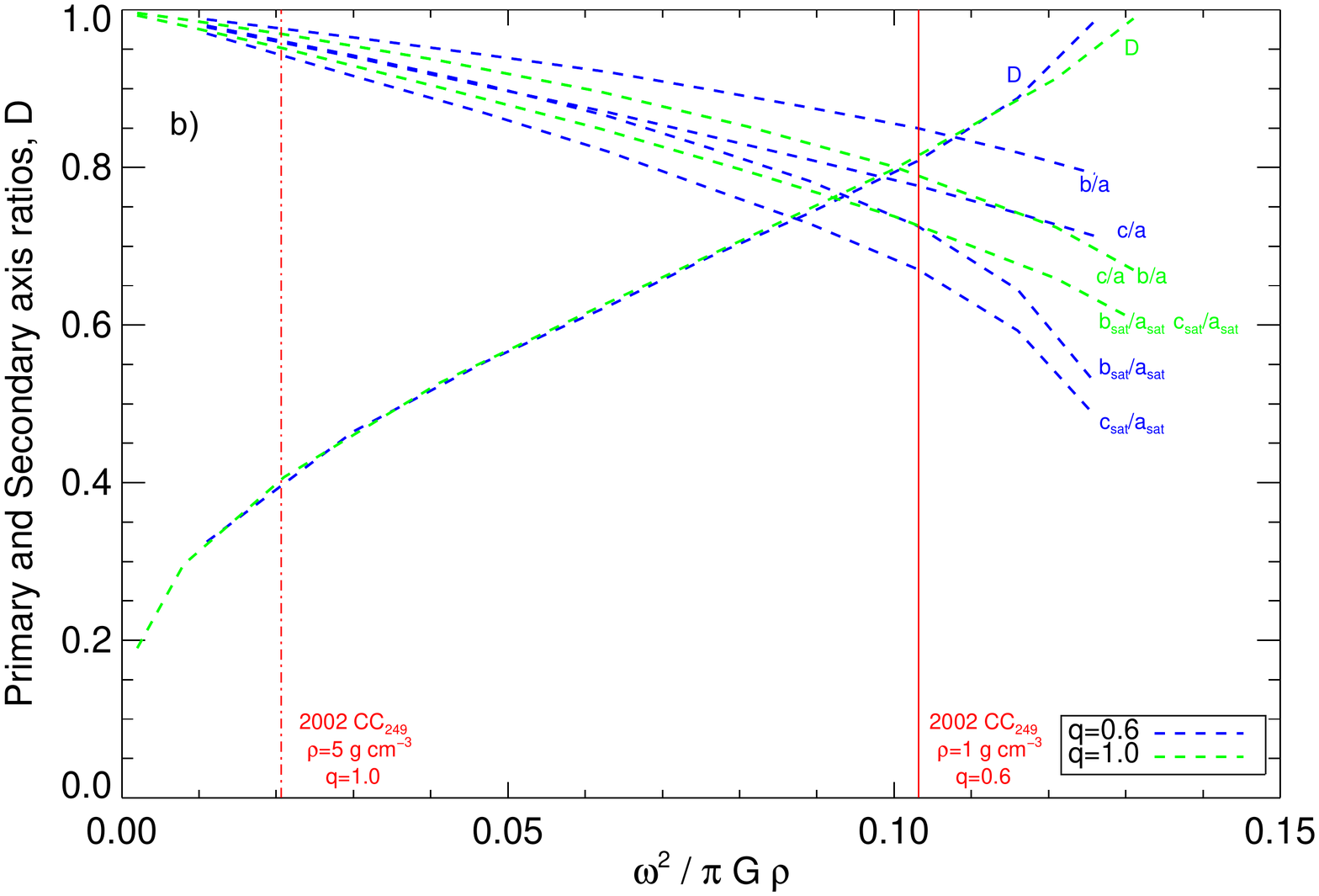}
\caption{\textit{The network of Roche sequences (plot a)), adapted from \citet{Leone1984}), and axis ratios of the components, and parameter D (plot b))}: 2002~CC$_{249}$ (red square) can have a mass ratio of 1, and a density of 5~g~cm$^{-3}$, or a mass ratio of 0.6 and a density of 1~g~cm$^{-3}$. Axis ratios of the primary (b/a, c/a), of the secondary (b$_{sat}$/a$_{sat}$, c$_{sat}$/a$_{sat}$), and parameter (D) for a mass ratio of 0.6 (blue), and 1 (green). The red dot-dash line is 2002~CC$_{249}$ assuming a mass ratio of 1, and the red continuous line is using a mass ratio of 0.6. See \citet{Leone1984, Thirouin2017} for more details about the network of Roche sequences. }
\label{fig:Roche}
\end{figure*}

Following \citet{Leone1984}, the mass ratio and the density of the system are estimated (Figure~\ref{fig:Roche}). Two extreme options (min and max) are obtained: i) a mass ratio q$_{min}$=0.6 and density $\rho_{max}$=1~g~cm$^{-3}$ or ii) a mass ratio of q$_{max}$=1 and density $\rho_{max}$=5~g~cm$^{-3}$. The uncertainty for the mass ratio is $\pm$0.05. For the rest of the study, conservative mass ratios of q$_{min}$=0.6, and q$_{max}$=1 will be used (reasons presented in \citet{Thirouin2017}).

If 2002~CC$_{249}$ is a Roche system with q=0.6, and $\rho$=1~g~cm$^{-3}$, the primary's axis ratios are: b/a=0.85, c/a=0.78 (a=125/55~km, b=106/47~km, and c=97/43~km considering a geometric albedo of 0.04/0.20, and H=6.15~mag), the secondary's axis ratios are: b$_{sat}$/a$_{sat}$=0.73, c$_{sat}$/a$_{sat}$=0.67 (a=117/52~km, b=85/38~km, and c=78/35~km with an albedo of 0.04/0.20). The value \footnote{D=(a+a$_{sat}$)/d with the orbital separation (d), and a, a$_{sat}$ the primary and secondary longest axes respectively} D is 0.81. Therefore, the separation between the components is 299/132~km considering an albedo of 0.04/0.20.

With q=1 and $\rho$=5~g~cm$^{-3}$, the axis ratios of the primary are: b/a=0.97, c/a=0.95, and the secondary's ones are: b$_{sat}$/a$_{sat}$=0.97, c$_{sat}$/a$_{sat}$=0.95, and D=0.41. 

A density of 5~g~cm$^{-3}$ is improbable for an object with a diameter in the 200-400~km range, and especially for an object at the edge of our Solar System. Therefore, the option considering $\rho$=1~g~cm$^{-3}$ is favored. But, only several lightcurves obtained at different system's geometries will be required to model the system and improve our estimates.  

\subsection{Jacobi ellipsoid}

A Fourier series (second order, generally able to reproduce lightcurves due to shape) failed to reproduce the lightcurve, since the lightcurve is not a simple sinusoid but has an U-/V-shape to it (Figure~\ref{fig:LCLomb}). This is why we prefer the contact binary hypothesis. 

However, following we also present a study assuming that 2002~CC$_{249}$ is a Jacobi ellipsoid.
 
Following \citet{Binzel1989}, the lightcurve amplitude ($\Delta$${m}$) of a Jacobi with a$>$b$>$c and in rotation along the c-axis varies as: 
\begin{equation}
\Delta m = 2.5 \log \left(\frac{a}{b}\right) - 1.25 \log \left(\frac{a^2 \cos ^2 \xi + c^2 \sin ^2 \xi}{b^2 \cos ^2 \xi + c^2 \sin ^2 \xi}\right)
\end{equation}
Considering an aspect angle ($\xi$) of 90$^\circ$, we estimate the object's elongation, a/b=2.07, and c/a=0.37 (c/a ratio estimated based on \citet{Chandrasekhar1987}). Therefore, using the previous axis ratio estimates and the absolute magnitude reported in this work, we compute: a=373/167~km, b=180/81~km, and c=138/62~km using 0.04/0.20 as albedo values and $\xi$=90$^\circ$.

With $\xi$=60$^\circ$, we derive an elongation larger than 2.31 indicating that the object is unstable to fission due to rotation \citep{Sheppard_phd, Jeans1919}. Therefore, if 2002~CC$_{249}$ is a Jacobi ellipsoid, its viewing angle has to be between 76$^\circ$ and 90$^\circ$.

Assuming an equatorial view and based on \citet{Chandrasekhar1987}, we compute the lower density limit: $\rho$$\geq$0.34~g~cm$^{-3}$. Such a low value favors an icy composition. This result is compatible with thermal modeling of TNOs from Herschel Space Observatory and/or Spitzer, suggesting a highly porous surface for these outer Solar System objects \citep{Lellouch2013, Vilenius2014}.


\section{Summary} 
 
Based on images carried out using the Lowell's Discovery Channel Telescope and the Magellan-Baade Telescope in 2016 and 2017, we summarize our results as follows: 

\begin{itemize}

\item 2002~CC$_{249}$ has an asymmetric double-peaked lightcurve with a U-/V-shape at the maximum/minimum of brightness, a periodicity of 11.87~h, and a peak-to-peak amplitude of 0.79~mag. This extreme variability is best interpreted by a contact binary. 2002~CC$_{249}$ is the first contact binary candidate in the dynamically Cold Classical population. This is surprising as the largest fraction of wide binaries is in this population.   

\item Assuming a contact binary, two main solutions are found: i) q$_{min}$=0.6, and $\rho_{min}$=1~g cm$^{-3}$ or ii) q$_{max}$=1, and $\rho_{max}$=5~g cm$^{-3}$. 

\item Because a density of $\rho$=5~g cm$^{-3}$ is doubtful for 2002~CC$_{249}$, we prefer the option with q=0.6, and $\rho$=1~g cm$^{-3}$. With this option, we find: b/a=0.85, c/a=0.78 for the primary, and b$_{sat}$/a$_{sat}$=0.73, c$_{sat}$/a$_{sat}$=0.67 for the secondary. We calculate that the components are separated by 299/132~km (using 0.04/0.20 as albedo range).

\item If 2002~CC$_{249}$ is a Jacobi in hydrostatic equilibrium, we estimate: $\rho$$\geq$0.34~g cm$^{-3}$, and a/b=2.07, assuming a viewing angle $\xi$=90$^\circ$. Its viewing angle must be between 76$^\circ$ and 90$^\circ$, if 2002~CC$_{249}$ is rotationally stable.

\item We report a new color study confirming that 2002~CC$_{249}$ has an ultra red surface, like most Cold Classical objects.

\item Contact binaries (likely/confirmed) present maxima of brightness (U-shape) with a larger full width at half maximum (FWHM), and smaller minima of brightness (V-shape) FWHM than single objects. The FWHM of the contact binaries U-shape is larger than 0.30, whereas other objects have a FWHM$\leqslant$0.28. The V-shape has a FWHM generally less than 0.21 for the (likely/confirmed) contact binaries. The FWHM difference in minimum and maximum peaks is greater than about 0.1 for contact binaries, and less than 0.05 for other objects, when the viewing angle is near equator-on. In the case of 2002~CC$_{249}$, the U-FWHM are 0.33 and 0.30 whereas the V-FHWM are 0.19 and 0.20 (one value per peak).

\end{itemize}

%
 
\acknowledgments

We would like to thank the referee for a careful reading of this paper and very useful comments. 
This research is based on data obtained at the Lowell Observatory's Discovery Channel Telescope (DCT). Lowell operates the DCT in partnership with Boston University, Northern Arizona University, the University of Maryland, and the University of Toledo. Partial support of the DCT was provided by Discovery Communications. LMI was built by Lowell Observatory using funds from the National Science Foundation (AST-1005313). We acknowledge the DCT operators: Andrew Hayslip, Heidi Larson, Teznie Pugh, and Jason Sanborn. A special thank to Jason for dealing with some technical difficulties during our DCT runs. This paper includes data gathered with the 6.5~m Magellan-Baade Telescope located at las Campanas Observatory, Chile. Audrey Thirouin is partly supported by Lowell Observatory funding. Authors acknowledge support from the National Science Foundation, grant number AST-1734484 awarded to the ``Comprehensive Study of the Most Pristine Objects Known in the Outer Solar System".

 \clearpage

\onecolumn

\begin{table}[h!]
\caption{\label{Tab:Log_Obs} UT-Dates (MM/DD/YYYY), the number of images (Nb.) obtained each night, the heliocentric (r$_{h}$), and geocentric ($\Delta$) distances in astronomical units (AU), the phase angle ($\alpha$, in degrees) of the observations, the filter(s) used, and the telescope are reported in this Table.} Geocentric, heliocentric distances and phase angle are from the Minor Planet Center Ephemeris generator.   
\center
\begin{tabular}{ccccccc} 

\hline
UT-date & Nb.   & r$_h$ &  $\Delta$ & $\alpha$   & Filter & Telescope\\
 &       &  [AU]  &  [AU]  &  [$^{\circ}$]   & & \\
\hline
\hline
02/14/2016  &  23   & 38.059   & 37.334  &  1.0    & VR & DCT\\
04/06/2016  &  19  &  38.056  & 37.066  &   0.2   & VR & DCT\\    
05/11/2016  &  1+1+1  &  38.054   & 37.312  & 1.0   & g'r'i' & DCT \\   
05/13/2016  &  16  &  38.054  & 37.334 &  1.1   & VR & DCT\\ 
03/10/2017  &  12  & 38.036  & 37.106  &  0.5   & r' & DCT\\ 
03/18/2017  &  38  & 38.036   & 37.064  &  0.3    & VR& DCT \\ 
03/19/2017  &  3  & 38.036   & 37.061  &  0.3   & VR & DCT\\ 
03/20/2017  &  20  & 38.036   & 37.056 &  0.3    & VR & DCT\\ 
03/30/2017  &  67  & 38.035   & 37.037  &  0.0    & VR & DCT \\ 
04/24/2017  &  2+2+2 &  38.034   &  37.116 & 0.6  & g'r'i' & Magellan \\ 
\hline
\hline

\end{tabular}
\end{table}

 \clearpage

 \clearpage

 \clearpage

 \clearpage

\begin{deluxetable}{lccc}
 \tablecaption{\label{Tab:Summary_photo} Photometry used in this paper is available in the following table. Julian date is without light-time correction. }
\tablewidth{0pt}
\tablehead{
Object  & Julian Date & Relative magnitude &  Error  \\
        &          &   [mag]           &  [mag]\\
}
\startdata  
2002~CC$_{249}$ &   &   &   \\
& 2457432.88848 & 0.07 & 0.07 \\ 
&2457432.89676 & -0.06 & 0.14 \\
& 2457432.90608 & -0.06 & 0.04 \\ 
&2457432.91436 & -0.14 & 0.03 \\ 
&2457432.92265 & -0.20 & 0.03 \\
& 2457432.92979 & -0.24 & 0.03 \\ 
&2457432.93689 & -0.28 & 0.03 \\ 
&2457432.94402 & -0.30 & 0.03 \\
& 2457432.95111 & -0.36 & 0.03 \\ 
&2457432.95824 & -0.35 & 0.03 \\
& 2457432.96535 & -0.35 & 0.03 \\
& 2457432.97247 & -0.38 & 0.03 \\ 
&2457432.97957 & -0.37 & 0.04 \\ 
&2457432.98669 & -0.32 & 0.04 \\
& 2457432.99382 & -0.30 & 0.05 \\ 
&2457433.00095 & -0.22 & 0.04 \\ 
&2457433.00922 & -0.19 & 0.04 \\ 
&2457433.01751 & -0.14 & 0.04 \\ 
&2457433.02579 & -0.11 & 0.04 \\
& 2457433.03407 & -0.09 & 0.05 \\ 
&2457433.04235 & -0.05 & 0.05 \\
 & 2457484.73942 & -0.02 & 0.11 \\ 
& 2457484.74778 & 0.01 & 0.16 \\
 & 2457484.75619 & 0.14 & 0.16 \\ 
& 2457484.76586 & 0.30 & 0.13 \\ 
& 2457484.77536 & 0.33 & 0.13 \\ 
& 2457484.78552 & 0.34 & 0.10 \\
 & 2457484.79505 & 0.34 & 0.10 \\
 & 2457484.80463 & 0.43 & 0.12 \\
 & 2457484.81407 & 0.28 & 0.17 \\ 
& 2457484.82355 & 0.13 & 0.19 \\ 
& 2457484.83300 & 0.04 & 0.13 \\ 
& 2457484.84244 & -0.01 & 0.08 \\ 
& 2457484.85189 & -0.11 & 0.05 \\
 & 2457484.86134 & -0.22 & 0.05 \\ 
& 2457484.87078 & -0.23 & 0.07 \\
 & 2457484.88023 & -0.22 & 0.06 \\ 
& 2457484.88968 & -0.39 & 0.05 \\ 
& 2457484.89912 & -0.35 & 0.08 \\
 & 2457484.90858 & -0.39 & 0.09 \\
 & 2457521.67318 & 0.23 & 0.04 \\
 & 2457521.68144 & 0.19 & 0.04 \\
 & 2457521.68972 & 0.08 & 0.04 \\ 
& 2457521.69799 & 0.02 & 0.03 \\ 
& 2457521.70626 & -0.03 & 0.03 \\ 
& 2457521.71453 & -0.13 & 0.03 \\ 
& 2457521.72279 & -0.15 & 0.03 \\
 & 2457521.73104 & -0.25 & 0.03 \\ 
& 2457521.73931 & -0.24 & 0.03 \\
 & 2457521.74772 & -0.30 & 0.03 \\
 & 2457521.75598 & -0.29 & 0.03 \\ 
& 2457521.76425 & -0.27 & 0.03 \\ 
& 2457521.77250 & -0.30 & 0.03 \\ 
& 2457521.78076 & -0.28 & 0.03 \\ 
& 2457521.78911 & -0.31 & 0.03 \\
 & 2457521.79737 & -0.29 & 0.03 \\
 & 2457822.85438 & -0.17 & 0.06 \\ 
& 2457822.86735 & -0.15 & 0.07 \\ 
& 2457822.88009 & -0.07 & 0.06 \\
 & 2457822.89282 & 0.20 & 0.09 \\ 
& 2457822.90551 & 0.27 & 0.09 \\ 
& 2457822.91818 & 0.37 & 0.09 \\
 & 2457822.93084 & 0.38 & 0.11 \\
 & 2457822.94353 & 0.38 & 0.13 \\
 & 2457822.95619 & 0.15 & 0.10 \\
 & 2457822.96892 & -0.01 & 0.11 \\
 & 2457822.98330 & -0.11 & 0.12 \\ 
& 2457822.99580 & -0.20 & 0.08 \\
 & 2457830.81664 & 0.20 & 0.05 \\
 & 2457830.82097 & 0.30 & 0.05 \\ 
& 2457830.83066 & 0.40 & 0.05 \\ 
& 2457830.83544 & 0.39 & 0.05 \\ 
& 2457830.84025 & 0.32 & 0.06 \\ 
& 2457830.84503 & 0.29 & 0.05 \\ 
& 2457830.85821 & 0.22 & 0.05 \\ 
& 2457830.86299 & 0.09 & 0.05 \\ 
& 2457830.86780 & 0.09 & 0.05 \\ 
& 2457830.87258 & 0.01 & 0.04 \\
 & 2457830.87740 & -0.05 & 0.04 \\
 & 2457830.88218 & -0.06 & 0.04 \\
 & 2457830.88699 & -0.13 & 0.05 \\ 
& 2457830.89178 & -0.13 & 0.04 \\ 
& 2457830.89659 & -0.13 & 0.04 \\
 & 2457830.89659 & -0.16 & 0.04 \\
 & 2457830.90138 & -0.18 & 0.04 \\
 & 2457830.90619 & -0.20 & 0.04 \\
 & 2457830.91097 & -0.22 & 0.04 \\
 & 2457830.91579 & -0.27 & 0.04 \\
 & 2457830.92057 & -0.31 & 0.04 \\ 
& 2457830.92538 & -0.29 & 0.04 \\ 
& 2457830.93017 & -0.30 & 0.05 \\
 & 2457830.93498 & -0.33 & 0.07 \\
 & 2457830.93977 & -0.33 & 0.07 \\
 & 2457830.94457 & -0.32 & 0.04 \\ 
& 2457830.94939 & -0.33 & 0.03 \\
 & 2457830.95420 & -0.34 & 0.04 \\ 
& 2457830.95900 & -0.30 & 0.05 \\
 & 2457830.96380 & -0.31 & 0.06 \\
 & 2457830.96861 & -0.32 & 0.07 \\ 
& 2457830.97339 & -0.31 & 0.07 \\
 & 2457830.97821 & -0.34 & 0.08 \\ 
& 2457830.98299 & -0.30 & 0.08 \\
 & 2457831.71589 & -0.22 & 0.03 \\
 & 2457831.74388 & -0.16 & 0.04 \\ 
& 2457831.91506 & -0.21 & 0.04 \\
 & 2457832.90838 & -0.23 & 0.03 \\ 
& 2457832.91319 & -0.28 & 0.04 \\ 
& 2457832.91797 & -0.29 & 0.05 \\ 
& 2457832.92279 & -0.32 & 0.06 \\ 
& 2457832.92758 & -0.30 & 0.07 \\ 
& 2457832.93238 & -0.29 & 0.07 \\ 
& 2457832.93718 & -0.29 & 0.08 \\ 
& 2457832.94198 & -0.28 & 0.08 \\ 
& 2457832.94677 & -0.28 & 0.03 \\ 
& 2457832.95159 & -0.29 & 0.04 \\ 
& 2457832.95637 & -0.32 & 0.05 \\ 
& 2457832.96118 & -0.30 & 0.06 \\ 
& 2457832.96596 & -0.32 & 0.07 \\ 
& 2457832.97078 & -0.32 & 0.07 \\ 
& 2457832.97557 & -0.34 & 0.08 \\
 & 2457832.98037 & -0.30 & 0.08 \\ 
& 2457832.98516 & -0.18 & 0.03 \\ 
& 2457832.98997 & -0.18 & 0.04 \\ 
& 2457832.99476 & -0.21 & 0.04 \\ 
& 2457842.66412 & -0.05 & 0.04 \\ 
& 2457842.66890 & -0.02 & 0.03 \\
 & 2457842.67372 & 0.04 & 0.03 \\
 & 2457842.67851 & 0.10 & 0.04 \\
 & 2457842.68331 & 0.15 & 0.04 \\ 
& 2457842.68810 & 0.21 & 0.04 \\ 
& 2457842.69288 & 0.26 & 0.04 \\
 & 2457842.69767 & 0.30 & 0.04 \\
 & 2457842.70247 & 0.32 & 0.04 \\ 
& 2457842.70727 & 0.38 & 0.04 \\ 
& 2457842.71206 & 0.44 & 0.04 \\
 & 2457842.71684 & 0.40 & 0.04 \\ 
& 2457842.72163 & 0.38 & 0.04 \\
 & 2457842.72642 & 0.37 & 0.03 \\
 & 2457842.73123 & 0.36 & 0.04 \\ 
& 2457842.73602 & 0.29 & 0.03 \\ 
& 2457842.74080 & 0.23 & 0.03 \\ 
& 2457842.74559 & 0.19 & 0.03 \\
 & 2457842.75038 & 0.10 & 0.03 \\ 
& 2457842.75519 & 0.05 & 0.03 \\ 
& 2457842.75998 & -0.01 & 0.02 \\
 & 2457842.76476 & -0.05 & 0.02 \\
 & 2457842.76955 & -0.07 & 0.02 \\
 & 2457842.77433 & -0.10 & 0.02 \\
 & 2457842.77914 & -0.13 & 0.02 \\ 
& 2457842.78394 & -0.19 & 0.02 \\ 
& 2457842.78872 & -0.20 & 0.02 \\
 & 2457842.79351 & -0.23 & 0.02 \\
 & 2457842.79829 & -0.24 & 0.02 \\
 & 2457842.80310 & -0.26 & 0.02 \\
 & 2457842.80789 & -0.28 & 0.02 \\
 & 2457842.81267 & -0.32 & 0.02 \\
 & 2457842.81747 & -0.31 & 0.02 \\
 & 2457842.82225 & -0.31 & 0.02 \\
 & 2457842.82706 & -0.30 & 0.02 \\
 & 2457842.83185 & -0.29 & 0.02 \\
 & 2457842.83663 & -0.30 & 0.02 \\ 
& 2457842.84142 & -0.31 & 0.02 \\ 
& 2457842.84620 & -0.32 & 0.02 \\
 & 2457842.85102 & -0.31 & 0.02 \\
 & 2457842.85581 & -0.31 & 0.02 \\ 
& 2457842.86059 & -0.31 & 0.02 \\ 
& 2457842.86538 & -0.29 & 0.02 \\
 & 2457842.87016 & -0.30 & 0.02 \\
 & 2457842.87498 & -0.26 & 0.02 \\
 & 2457842.87977 & -0.22 & 0.02 \\
 & 2457842.88455 & -0.20 & 0.02 \\
 & 2457842.88934 & -0.18 & 0.02 \\ 
& 2457842.89412 & -0.14 & 0.02 \\
 & 2457842.89894 & -0.11 & 0.02 \\
 & 2457842.90373 & -0.09 & 0.02 \\ 
& 2457842.90851 & -0.07 & 0.02 \\
 & 2457842.91330 & 0.00 & 0.02 \\
 & 2457842.91808 & 0.03 & 0.03 \\
 & 2457842.92289 & 0.07 & 0.03 \\ 
& 2457842.92767 & 0.15 & 0.03 \\ 
& 2457842.93247 & 0.18 & 0.03 \\ 
& 2457842.93726 & 0.22 & 0.03 \\ 
& 2457842.94204 & 0.26 & 0.04 \\
 & 2457842.94685 & 0.32 & 0.04 \\
 & 2457842.95163 & 0.32 & 0.03 \\
 & 2457842.95642 & 0.33 & 0.04 \\ 
& 2457842.96122 & 0.33 & 0.04 \\ 
& 2457842.96600 & 0.39 & 0.04 \\ 
& 2457842.97081 & 0.36 & 0.05 \\
 & 2457842.97559 & 0.37 & 0.05 \\
 & 2457842.98041 & 0.32 & 0.05 \\ 
\enddata

\end{deluxetable}

\end{document}